\input{aipcheck}

\documentclass[
    ,final            
  ]
  {aipproc}

\layoutstyle{8x11single}

\def\ereco{E_{\rm rec}}
\begin{document}

\title{ CCQE, 2p2h excitations and $\nu-$energy reconstruction}


\author{J. Nieves}{
  address={Instituto de F\'\i sica Corpuscular (IFIC), Centro Mixto
Universidad de Valencia-CSIC, Institutos de Investigaci\'on de
Paterna, E-46071 Valencia, Spain}
}

\author{I. Ruiz Simo}{
  address={Dipartimento di Fisica, Universit\`a di Trento, I-38123 Trento, Italy}
}

\author{F. S\'anchez}{
  address={Institut de Fisica d$'$Altes Energies (IFAE), Bellaterra
  Barcelona, Spain}
}

\author{M. J. Vicente Vacas}{
  address={Departamento de F\'\i sica Te\'orica and IFIC, Centro Mixto
Universidad de Valencia-CSIC, Institutos de Investigaci\'on de
Paterna, E-46071 Valencia, Spain}
}

\begin{abstract}
 We analyze the MiniBooNE muon neutrino CCQE-like $d\sigma/dT_\mu d\cos\theta_\mu$
 data using a theoretical model that, among other nuclear effects,
 includes RPA correlations and 2p2h (multinucleon) mechanisms.  These
 corrections turn out to be essential for the description
 of the  data.  We find that MiniBooNE CCQE-like data are
 fully compatible with former determinations of the nucleon axial mass
 $M_A\sim 1.05$ GeV. This is in sharp contrast with several previous
 analysis where anomalously large values of $M_A\sim 1.4$ GeV have
 been suggested. We also show that because of the the multinucleon
 mechanism effects, the algorithm used to reconstruct the neutrino
 energy is not adequate when dealing with quasielastic-like events.
 Finally, we analyze the MiniBooNE unfolded cross section, and show
 that it exhibits an excess (deficit) of low (high) energy neutrinos,
 which is an artifact of the unfolding process that ignores 2p2h
 mechanisms.

\end{abstract}

\maketitle


\section{Introduction}

A correct understanding of neutrino-nucleus interactions is crucial to
minimize systematic uncertainties in neutrino oscillation
experiments \cite{FernandezMartinez:2010dm}.  Most of the new
generation of neutrino experiments are exploring neutrino-nuclear
scattering processes at intermediate energies ($\le 2$ GeV), thus
experiments like ScibooNE \cite{Nakajima:2010fp} or MiniBooNE \cite{AguilarArevalo:2010zc,
AguilarArevalo:2010bm,AguilarArevalo:2010xt,AguilarArevalo:2013hm} have produced good quality data for
quasi-elastic scattering and pion production in this neutrino energy
region. These new data show interesting deviations from the
predictions of present models that have raised doubts in the areas which seemed to
be well understood \cite{Morfin:2012kn,Gallagher:2011zza}.  
The list of new puzzles is quite long and seems to
be expanding. In this talk, we focus in
particular on charged-current quasi-elastic (CCQE) scattering, and we
would try to shed some light into three of these puzzles: 
i) What is the value of the nucleon axial mass? ii) How
large is the two-body current contribution that can mimic genuine
QE interactions?, and iii)  What is the impact of the
 multinucleon processes on the neutrino energy reconstruction and on
 the neutrino flux-unfolded cross sections?

\section{CCQE-like scattering}
\begin{figure}
  \includegraphics[height=.3\textheight]{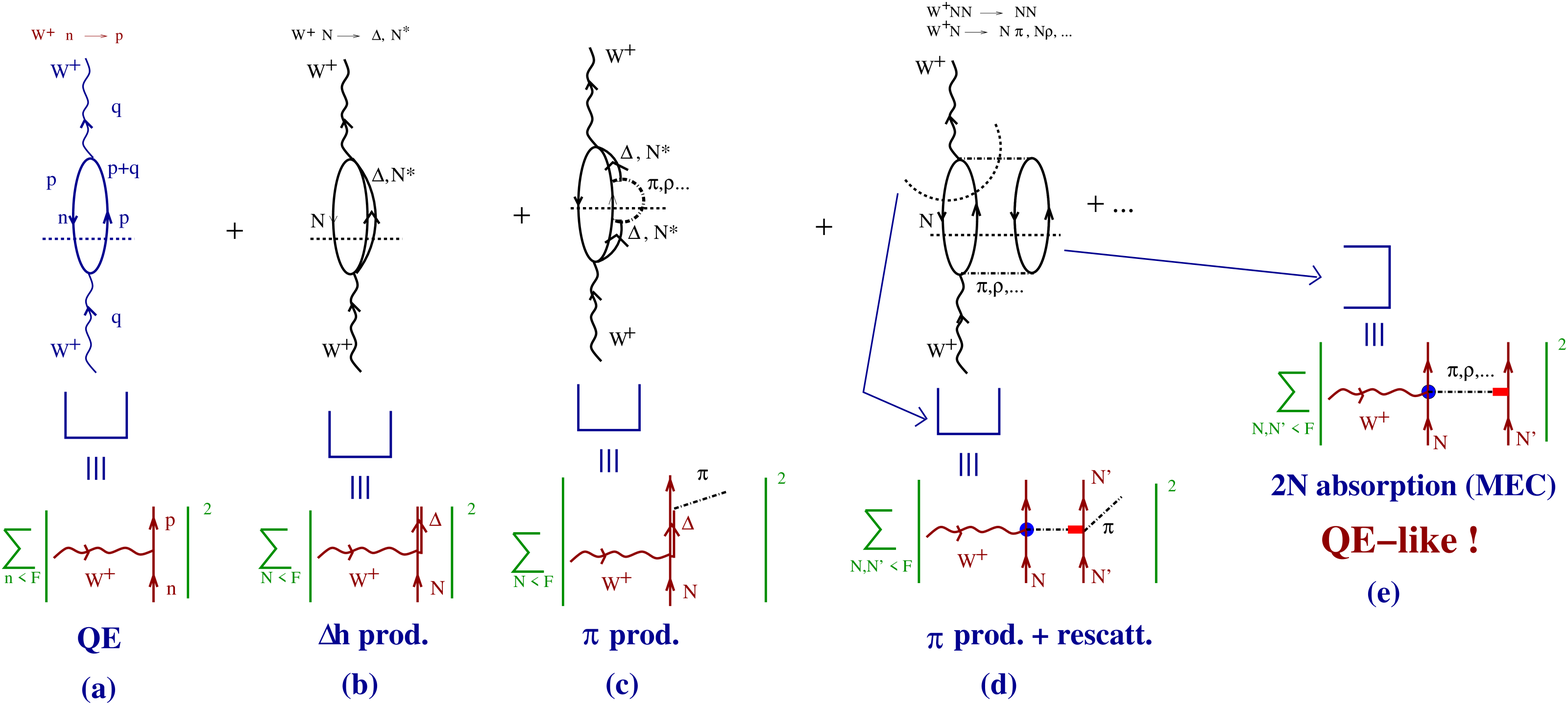}
  \caption{Diagrammatic representation of some diagrams
  contributing to the $W-$selfenergy and their connection with
  different absorption modes of the gauge boson in the nuclear
  medium. }\label{fig:fig2}
\end{figure}

The  inclusive cross section for the process $\nu_\ell (k) +\,
A_Z \to \ell^- (k^\prime) + X $ is determined by the $W$ gauge boson selfenergy in the nuclear
medium \cite{Nieves:2004wx,Nieves:2011pp}, and in particular for the
different modes in which it can be absorbed.  The most relevant ones
are: the absorption by one nucleon, or by a pair of correlated
nucleons that are exchanging virtual mesons ($\pi$, $\rho$, $\cdots$),
or the excitation of a $\Delta$ or a higher energy resonance,
etc. (see Fig.~\ref{fig:fig2}).  In most theoretical works QE is used
for processes where the gauge boson $W$ is absorbed by just one
nucleon, which together with a lepton is emitted (see
Fig.~\ref{fig:fig2}a). In what follows, we will refer to this
contribution as {\it genuine} QE. However, the recent MiniBooNE CCQE
data \cite{AguilarArevalo:2010zc} include events in which only a muon
is detected (we will refer to them as QE-like events). This data
selection is adopted because ejected nucleons are not detected in that
experiment.  Thus,  the QE-like sample does not include events
with pions coming off the nucleus, since they will give rise to
additional leptons after their decay (see Fig.~\ref{fig:fig2}c).
However, this event-sample includes multinucleon events, as those
displayed in Fig.~\ref{fig:fig2}e, where the gauge boson is absorbed
by two interacting nucleons (in the many body language, this amounts
to the excitation of a 2p2h nuclear component). On the other hand, other events like real
pion production followed by its absorption should be also included in
the QE-like sample, though the MiniBooNE analysis Monte Carlo corrects
for those. Here, there is a subtlety that is worth to comment in some
detail. Let us pay attention to processes like the one depicted in the bottom panel of
Fig.~\ref{fig:fig2}c, but when the pion is off-shell instead of being
on the mass-shell. In any of these processes, the \underline{virtual} pion, 
that is produced in the first step, will   be necessarily absorbed by a second
nucleon, and thus the process should be classified/cataloged as a two nucleon $W$ absorption
mechanism (Fig.~\ref{fig:fig2}e).  Hence, events originated by these kind of processes  
do not contribute to the genuine QE
cross section, but they do to the cross section measured in the MiniBooNE
experiment.

After this discussion, we draw a first important conclusion: the
MiniBooNE CCQE data \cite{AguilarArevalo:2010zc} cannot be directly
compared to most of the previous theoretical calculations, in which
only the one-body genuine QE contribution was usually considered. This
was first pointed out by M. Martini et
al.~\cite{Martini:2009uj,Martini:2010ex}. Indeed, the absolute values
of the CCQE cross section reported in \cite{AguilarArevalo:2010zc} are
too large as compared to the consensus of theoretical predictions for
the genuine QE contribution \cite{Boyd:2009zz}. Thus, the cross
section per nucleon on $^{12}$C is clearly larger than for free
nucleons, and a fit, using a relativistic Fermi gas model, to the data
led to an axial mass, $M_A=1.35\pm 0.17$
GeV \cite{AguilarArevalo:2010zc} , much larger than the previous world
average ($\approx 1.03$ GeV).  Similar results 
have been later obtained analyzing MiniBooNE data with
more sophisticated treatments of the nuclear effects that work well in
the study of electron scattering \cite{Benhar:2010nx,
Juszczak:2010ve,Butkevich:2010cr}. 

In what follows, we present results from a microscopic
calculation \cite{Nieves:2004wx, Nieves:2011pp} of the CCQE-like 2D
cross section $d\sigma/dT_\mu d\cos\theta_\mu$. There are no free
parameters in the description of nuclear effects, since they were
fixed in previous studies of photon, electron, and pion interactions
with nuclei \cite{Carrasco:1989vq,Nieves:1993ev,Nieves:1991ye,Gil:1997bm}.
 We approximate the CCQE-like cross section by the sum of the
genuine QE  contribution (Fig.~\ref{fig:fig2}a) and that induced by
2p2h mechanisms (Fig.~\ref{fig:fig2}e), for which the gauge boson is
being absorbed by two or more nucleons without producing pions.

\begin{figure}[tbh]
\includegraphics[height=6.0cm]{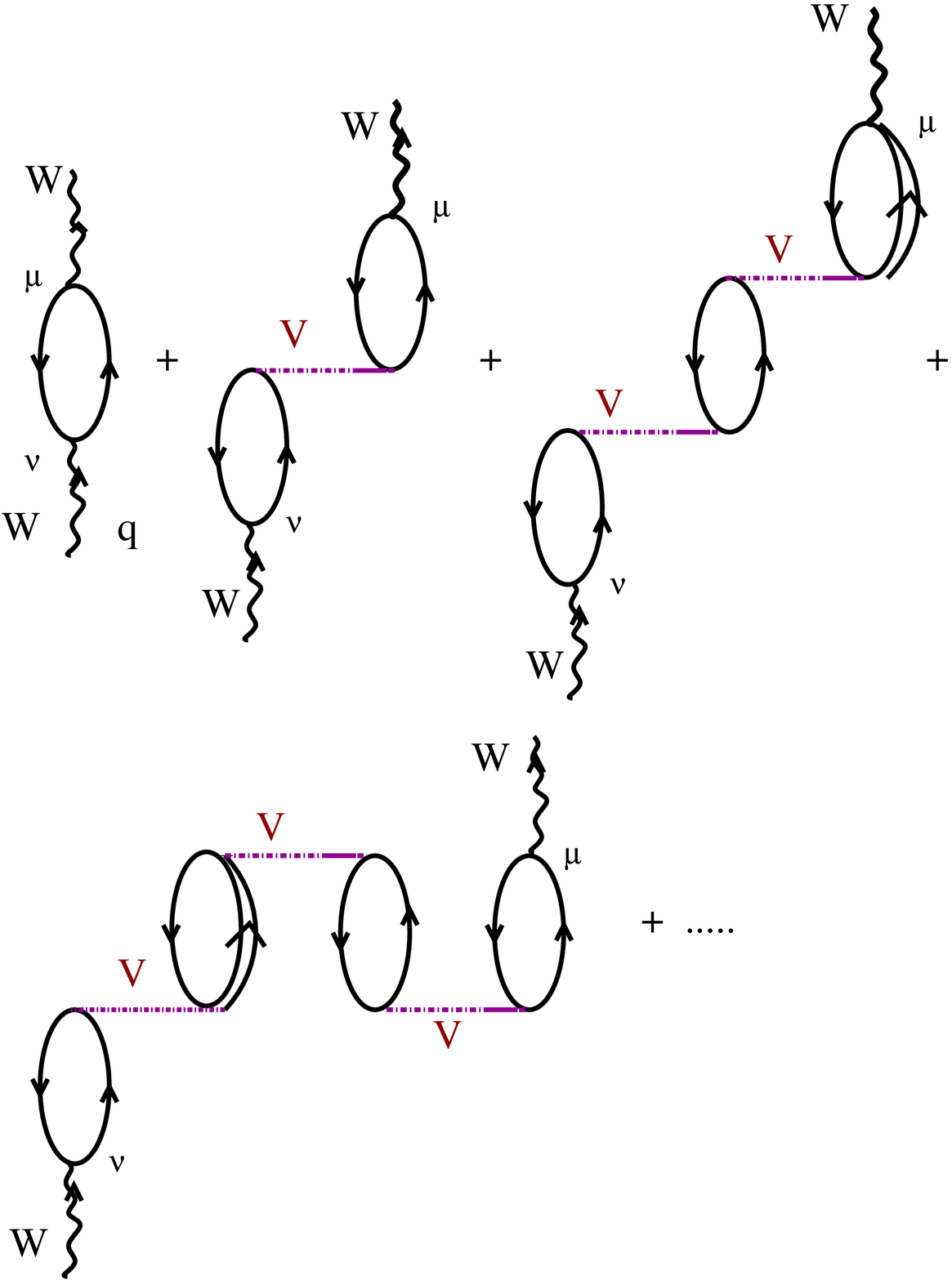}\hspace{1cm}\includegraphics[width=0.35\textwidth]{crosssectionunfolded.final.eps}
\caption{Left: Set of irreducible diagrams responsible
  for the polarization (RPA) effects in the 1p1h contribution to the
  $W$ self-energy.  Right. Theoretical $\sigma$ and approximate $\sigma_{\rm appx}$ 
CCQE-like integrated cross sections in carbon as a
function of the neutrino energy. Results have been obtained from
Refs. \cite{Nieves:2004wx} and \cite{Nieves:2011pp} using $M_A \sim 1.05$ GeV. 
Details can be found in Ref. \cite{Nieves:2012yz}.   The MiniBooNE
  data~\cite{AguilarArevalo:2010zc} and errors (shape) have been re-scaled by
a factor 0.89. }\label{fig:fig3}
\end{figure}

\begin{figure}[tbh]
\hspace{-0.85cm}\includegraphics[height=3.3cm]{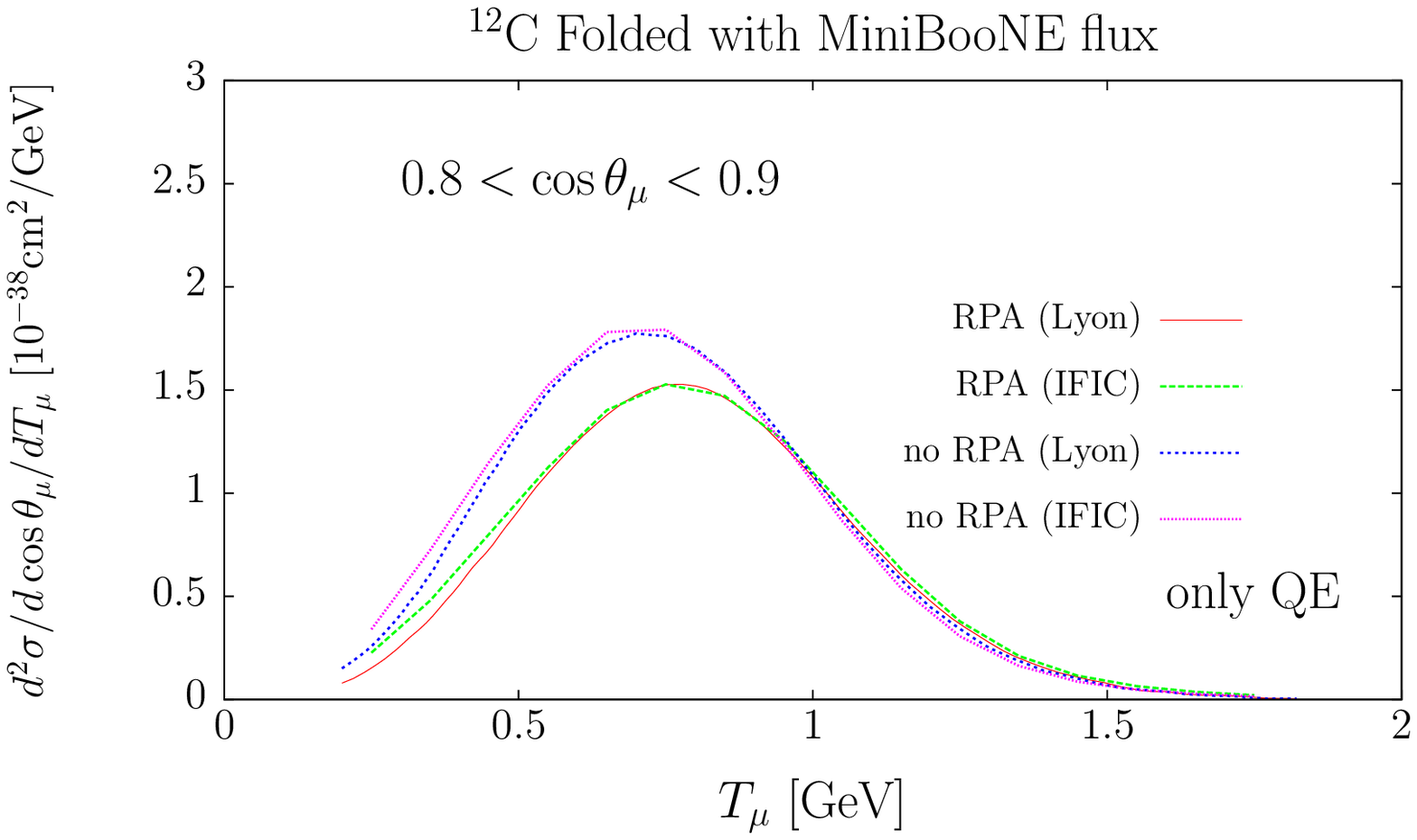}\hspace{-0.8cm}\includegraphics[height=3.7cm]{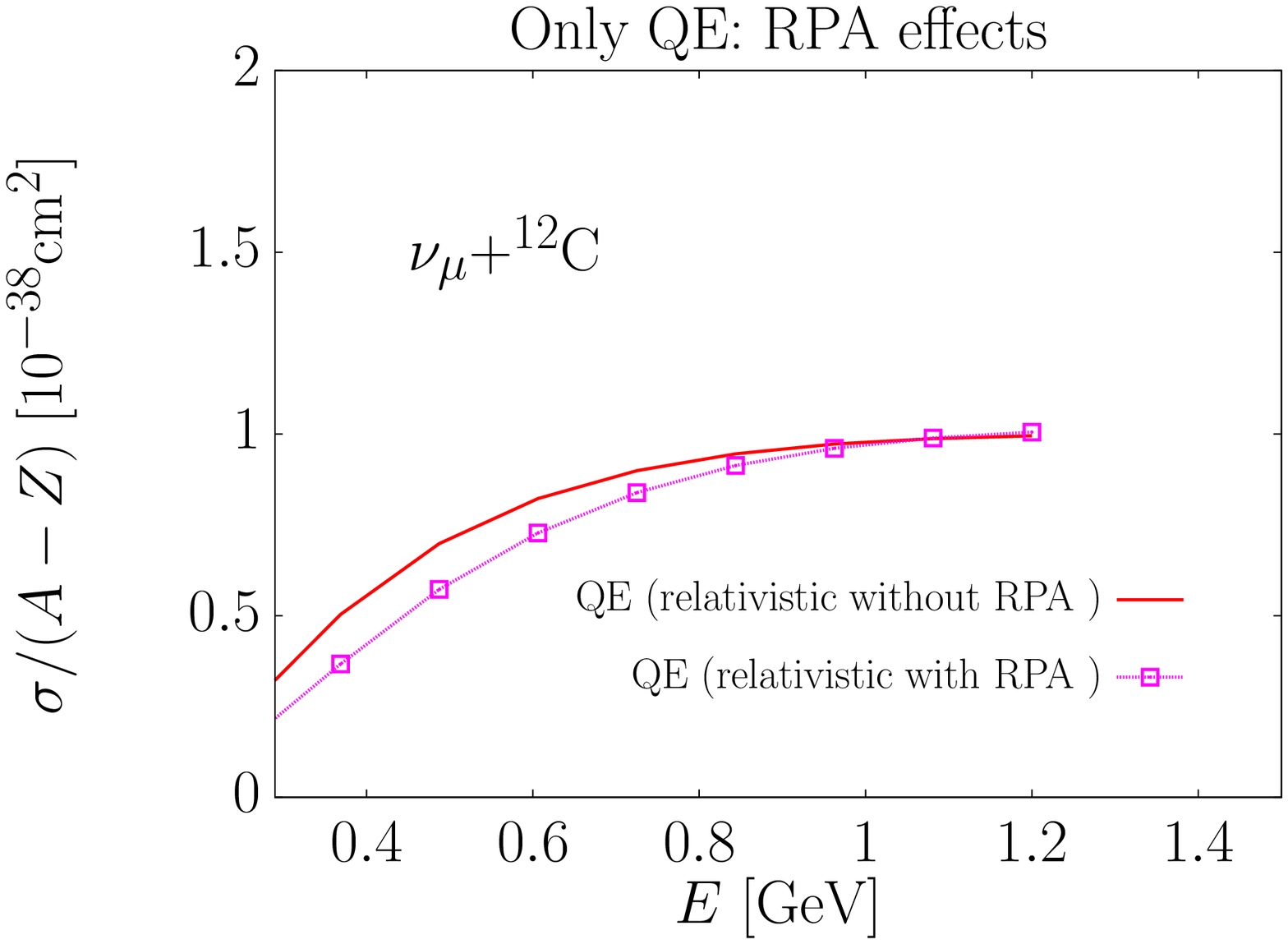}\hspace{-0.8cm}\includegraphics[height=3.7cm]{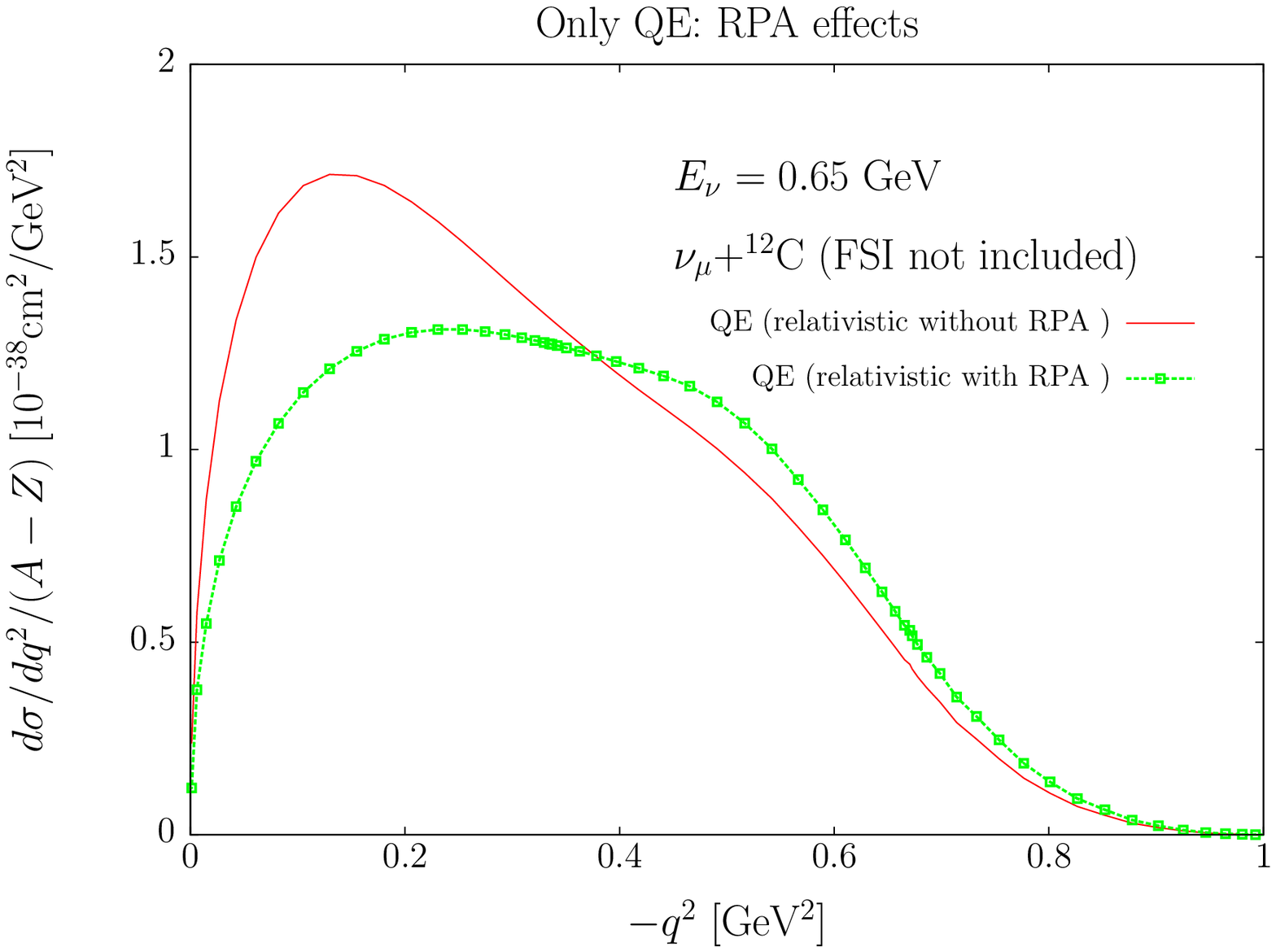}
\caption{Right: MiniBooNE flux-averaged  $\nu_\mu- ^{12}$C double differential cross section per
  neutron for $0.8 < \cos\theta_\mu < 0.9$ as a function of the muon
  kinetic energy. The other two plots correspond to different theoretical predictions for muon
  neutrino CCQE total cross section off $^{12}$C as a function of the
  neutrino energy (middle) and $q^2$ (right), obtained from the relativistic model of
  Ref.~\cite{Nieves:2004wx}.  In all cases $M_A \sim 1.05$ GeV. }\label{fig:fig3bis}
\end{figure}

The genuine QE contribution was studied in \cite{Nieves:2004wx}
incorporating several nuclear effects. The main one is the medium
polarization (RPA), that accounts for the change of the electroweak
coupling strengths, from their free nucleon values, due to the
presence of strongly interacting nucleons.  Indeed, the quenching of
axial current is a well-established phenomenon. The RPA re-summation
accounts for the medium polarization effects in the 1p1h contribution
(Fig. \ref{fig:fig2}(a)) to the $W$ selfenergy by substituting it by a
collective response as shown diagrammatically in the left panel of
Fig. \ref{fig:fig3}. Evaluating these effects, requires of an in
medium baryon-baryon effective force, that within our model includes
$\Delta$-hole degrees of freedom, short range correlations and explicit $\pi$ and $\rho$ meson
exchanges in the vector-isovector channel. RPA effects are important,
as can be appreciated in Fig. \ref{fig:fig3bis}. In the left panel, we show results \cite{Nieves:2011yp} for the genuine QE
contribution from our model (labeled as IFIC)  for the CC quasielastic $\nu_\mu- ^{12}$C
double differential cross sections convoluted with the MiniBooNE
flux. There, we also display results from the model of M. Martini et
al. (labeled as Lyon) taken from \cite{Martini:2011wp}.  The
predictions of both groups for this genuine QE contribution, with and
without RPA effects, turn out to be in a quite good agreement.  
We would finally like to remark that the RPA corrections strongly decrease as the neutrino
energy increases, while they strongly modify the
$q^2-$differential distributions at low neutrino energies, as can be appreciated in the middle and right panels of
Fig. \ref{fig:fig3bis}, respectively.

The model for multinucleon mechanisms is fully
microscopical and it is discussed in detail 
in \cite{Nieves:2011pp}. It includes one, two, and even
three-nucleon mechanisms, as well as the excitation of $\Delta$
isobars.  This theoretical model has proved to be quite successful in the study
of nuclear reactions with photon \cite{Carrasco:1989vq},
pion \cite{Nieves:1993ev,Nieves:1991ye} and electron \cite{Gil:1997bm}
probes.

Up to neutrino energies around 1 GeV, the predictions of our model
obtained with $M_A=1.05$ GeV compare rather well \cite{Nieves:2011pp},
taking into account experimental and theoretical uncertainties, with
the data published by the SciBooNE collaboration for total neutrino
inclusive cross sections \cite{Nakajima:2010fp}. On the other hand,
the 2p2h contributions allows to describe \cite{Nieves:2011yp} the
CCQE-like flux averaged double differential 2D cross section
$d\sigma/dE_\mu d\cos\theta_\mu$ measured by MiniBooNE with values of
$M_A$ around $1.03\pm 0.02$ GeV that is usually quoted as the world
average. This is re-assuring from the theoretical point of view and
more satisfactory than the situation envisaged by some other works
that described these CCQE-like data in terms of a larger value of
$M_A$ of around 1.3--1.4 GeV, as mentioned above. The relative sizes
of the genuine QE and 2p2h cross sections in carbon can be appreciated
in the right panel of Fig. \ref{fig:fig3} by looking there at the
curves labeled as $\sigma^{ \rm QE (rel+RPA)}$ and $\sigma^{ \rm
2p2h}$, respectively.

The work of
Ref. \cite{Martini:2011wp} also include multinucleon
mechanisms and find a good description of the 2D MiniBooNE data. Both
works also agree on the relevant role played by the 2p2h mechanisms to
describe the MiniBooNE data.  However, both groups differ considerably
in the size (about a factor of two) of the multinucleon effects. There
exist indeed some important differences which amount to a more
comprehensive inclusion of mechanisms in our scheme and some
approximations used in the calculations of \cite{Martini:2011wp}. A
more detailed discussion on these differences can be found in
Refs.~\cite{Nieves:2011pp,Martini:2011ui}. We would also like to point out that the simple phenomenological
approach adopted in \cite{Lalakulich:2012ac} to account for the 2p2h
effects also reinforces the picture that emerges from 
\cite{Nieves:2011yp,Martini:2011wp}. Yet, a partial
microscopical calculation of the 2p2h contributions to the CCQE cross
section has been also presented in Refs.~\cite{Amaro:2010sd} and
\cite{Amaro:2011aa}, for neutrino and antineutrino induced reactions,
respectively. In these works, the contribution of the vector meson
exchange currents in the 2p2h sector is added to the QE neutrino or
antineutrino cross section predictions deduced from a phenomenological
model based on the super-scaling behavior
of electron scattering data.  In \cite{Amaro:2011qb}, and for the
neutrino case, the SuSA+2p2h results were also compared with those
obtained from a relativistic mean field approach. Although, all these
schemes do not account for the axial part of the 2p2h effects yet,
their results also corroborate that 2p2h meson exchange currents
play an important role in both CCQE-like neutrino and antineutrino
scattering, and that they may help to resolve the controversy on the
nucleon axial mass raised by the recent MiniBooNE data. 

A final remark concerns to the importance of 2p2h effects in
antineutrino reactions as compared to neutrino ones.  In our model the
relative importance of the 2p2h channel is somehow larger for
antineutrinos \cite{Nieves:2013fr}.  A similar trend, although with a
stronger reduction, has been found by Amaro et al.~\cite{Amaro:2011aa}
in the SuSA approximation.  However, other works like
Ref.~\cite{Bodek:2011ps}, which reaches agreement with MiniBooNE QE
neutrino data by modifying the magnetic form factors of the bound
nucleons, and Ref.~\cite{Martini:2010ex} lead to an enhancement of the
effect for antineutrino induced reactions.

\section{Neutrino Energy Reconstruction and the Shape of the
  CCQE-like Total Cross Section}

The relevance of the multinucleon mechanisms has some unwanted
consequences. Obviously, the neutrino energy reconstruction, based on
the QE kinematics is not so
reliable~\cite{Nieves:2012yz,Martini:2012fa,Lalakulich:2012hs,Martini:2012uc}
and that implies larger systematic uncertainties in the neutrino
oscillation experiments analysis. In general, the energy of the neutrino that has
originated an event is unknown, and  it is common to define a 
reconstructed neutrino energy $\ereco$, obtained from the measured
angle ($\theta_\ell$) and three-momentum ($\vec{p}_\ell$)
of the outgoing charged lepton $\ell$, as
\begin{equation}
\ereco = \frac{M
  E_\ell-m_\ell^2/2}{M-E_\ell+|\vec{p}_\ell|\cos\theta_\ell}\label{eq:defereco}
\end{equation}
which will correspond to the energy of a neutrino that emits a lepton,
of mass $m_\ell$ and energy, and a gauge
boson $W$ that is being absorbed by a nucleon of mass $M$ at rest. The
usual reconstruction procedure assumes that we are dealing with a
genuine quasielastic event on a nucleon at rest.

 Each event contributing to the flux
averaged double differential cross section $d\sigma/dE_\ell
d\cos\theta_\ell$ defines unambiguously a value of $\ereco$.  The
actual (``true'') energy, $E$, of the neutrino that has produced the
event will not be exactly $\ereco$.  Actually, for each $\ereco$,
there exists a distribution of true neutrino energies that could give
rise to events whose muon kinematics would lead to the given value of
$\ereco$. In the case of genuine QE events, this distribution is
sufficiently peaked  around the
true neutrino energy to make the algorithm in 
Eq.~(\ref{eq:defereco}) accurate enough to study the neutrino
oscillation phenomenon~\cite{Meloni:2012fq} or to extract neutrino
flux unfolded CCQE cross sections from data (assuming that  the neutrino flux
spectrum is known)~\cite{Nieves:2012yz,Martini:2012fa}. 

However, and due to the large importance of the 2p2h events, in
the case of CCQE-like events, there are appear a long tail in the
distribution of true energies associated to each $\ereco$ that makes
unreliable the use of Eq.~(\ref{eq:defereco}). The effects of the inclusion of multinucleon
processes on the energy reconstruction have been investigated within
our model in \cite{Nieves:2012yz}, finding results in a qualitative
agreement with those described in \cite{Martini:2012fa}.  In \cite{Nieves:2012yz}, it is also studied in detail the
$^{12}$C unfolded neutrino CCQE-like cross section published in
\cite{AguilarArevalo:2010zc}. Indeed, it is shown there, that it is
not a very clean observable, because the unfolding procedure itself is
model dependent and  assumes that the events are purely
QE. Moreover, it is also shown the MiniBooNE published cross section
differs from the real one $\sigma(E)$. This is illustrated in
the right panel of Fig.~\ref{fig:fig3}, where different predictions from our
model, together with the CCQE-like MiniBooNE data are depicted. The
theoretical results are obtained from the relativistic models of
Refs. \cite{Nieves:2004wx} and \cite{Nieves:2011pp}, for the genuine
QE and multinucleon contributions, respectively. In al cases $M_A$ is
set to 1.05 GeV. First, we see that the theoretical prediction
$\sigma^{\rm QE+\rm 2p2h}$ does not correctly reproduce the
neutrino-energy shape of the published data.  The 2p2h contributions
clearly improve the description of the data, which are totally missed
by the QE prediction. Though the model provides a reasonable
description, we observe a sizable excess of low energy neutrinos in
the data. The unfolding procedure (see Ref.~\cite{Nieves:2012yz} for
some details) does not appreciably distort the genuine QE events, and as can be
appreciated in the right panel of Fig.~\ref{fig:fig3}, $\sigma_{\rm appx}(E)$ is an
excellent approximation to the real $\sigma (E)$ cross section in that
case. However, the situation is drastically different for the 2p2h
contribution. It turns out that $\sigma_{\rm
  appx}^{\rm 2p2h}(E)$ (result obtained after the unfolding procedure)
is a poor estimate of the actual multinucleon mechanism contribution
$\sigma^{\rm 2p2h}(E)$. We also observe in Fig.~\ref{fig:fig3}
that the MiniBooNE CCQE-like data compare rather well with the
$\sigma_{\rm appx}$, quantity obtained after implementing the
unfolding procedure presumably carried out in
\cite{AguilarArevalo:2010zc}, but that however, appreciably differs
from the actual cross section $\sigma$.  Therefore, we conclude the MiniBooNE unfolded cross section
exhibits an excess (deficit) of low (high) energy neutrinos, which is
an artifact of the unfolding process that ignores multinucleon
mechanisms.

Similar conclusions are achieved in \cite{Nieves:2013fr}, where the
recent MiniBooNE antineutrino CCQE-like
data \cite{AguilarArevalo:2013hm} were discussed within our
framework. We show first that the model of
Refs. \cite{Nieves:2004wx,Nieves:2011pp}, that includes RPA and 2p2h
effects, satisfactorily describes the 2D data, and second that similar
limitations related to the energy reconstruction and the unfolding
procedure, when 2p2h effects are ignored, appear also for 
antineutrino CCQE processes.

%


\begin{theacknowledgments}
This research was supported by
the Spanish Ministerio de Econom\'\i a y Competitividad
and European FEDER funds under the contracts FIS2011-28853-C02-01,
FIS2011-28853-C02-02, FIS2011-24149,
 FPA2011-29823-C02-02 and the Spanish Consolider-Ingenio 2010
 Programme CPAN (CSD2007-00042),
by Generalitat Valenciana under contract PROMETEO/2009/0090 and by the 
EU Hadron-Physics2 project, grant agreement no. 227431.
\end{theacknowledgments}



\bibliographystyle{aipproc}   

\bibliography{nuint}

\IfFileExists{\jobname.bbl}{}
 {\typeout{}
  \typeout{******************************************}
  \typeout{** Please run "bibtex \jobname" to optain}
  \typeout{** the bibliography and then re-run LaTeX}
  \typeout{** twice to fix the references!}
  \typeout{******************************************}
  \typeout{}
 }

\end{document}